\documentclass[12pt,preprint]{aastex63}
\usepackage{epstopdf}
\usepackage{natbib}
\usepackage{multirow}
\usepackage{booktabs}
\usepackage{CJK}
\usepackage{hyperref}
\usepackage[figuresright]{rotating}

\submitjournal{AJ}

\begin{document}
\begin{CJK*}{UTF8}{gbsn}
\title{The Dust Mass of Supernova Remnants in M31 }

\author{Ye Wang(王野)}
\affiliation{Department of Astronomy,
	Beijing Normal University,
	Beijing 100875, China}

\author[0000-0003-3168-2617]{Biwei Jiang (姜碧沩)}
\affiliation{Department of Astronomy,
	Beijing Normal University,
	Beijing 100875, China}

\author[0000-0001-9328-4302]{Jun Li(李军)}
\affiliation{Department of Astronomy,
	Beijing Normal University,
	Beijing 100875, China}

\author[0000-0003-2645-6869]{He Zhao(赵赫)}
\affiliation{Department of Astronomy,
	Beijing Normal University,
	Beijing 100875, China}

\author[0000-0003-1218-8699]{Yi Ren(任逸)}
\affiliation{Department of Astronomy,
	Beijing Normal University,
	Beijing 100875, China}

\correspondingauthor{Biwei Jiang}
\email{bjiang@bnu.edu.cn}

\begin{abstract}

The dust temperature and mass of the supernova remnants (SNRs) in M31 are estimated by fitting the infrared spectral energy distribution calculated from the images in the $Spitzer$/IRAC4 and MIPS24, $Herschel$/PACS70, 100, 160, and $Herschel$/SPIRE250, 350\,$\mu$m band. Twenty SNRs with relatively reliable photometry exhibit an average dust temperature of $20.1^{+1.8}_{-1.5}$\,K, which is higher than the surrounding and indicating the heating effect of supernova explosion. The dust mass of these SNRs ranges from about 100 to 800\,$ M_{\odot}$, much bigger than the SNRs in the Milky Way. On the other hand, this yields the dust surface density of $0.10^{+0.07}_{-0.04}$\,${ M_{\odot} \rm pc^{-2}}$, about half of the surrounding area, which implies that about half dust in the SNRs is destroyed by the supernova explosion. The dust temperature, the radius, and thus the dust mass all demonstrate that the studied SNRs are old and very likely in the snowplow or even fade away phase because of the limitation by the far distance and observation resolution of M31, and the results can serve as a reference to the final effect of supernova explosion on the surrounding dust.

\end{abstract}


\section{Introduction}

In this paper we estimate the extent to which dust swept up by old supernova remnants in M31 has been destroyed by the associated shocks. The dust mass of the SNRs in M31 is estimated by fitting the spectral energy distribution composed of the fluxes measured from the images taken by \emph{Spitzer}/IRAC, MIPS and \emph{Herschel}/ PACS. The derived dust mass of SNRs is compared with the surrounding medium to discuss the amount of dust destroyed by the SN explosion. 

Dust is destroyed in SNR shocks primarily by sputtering and shattering. Sputtering is a process in which energetic particles knock atoms off the surface of the grain \citep{1996ApJ...457..244D}, more often in fast shocks with $v >$150\,km/s. It results in a deficit of small grains in SNRs. Shattering is a kind of grain-grain collisions, dominant in slower shocks with 50\,km/s$<v \leq$80\,km/s \citep{1994ApJ...433..797J}. Shattering is most effective on small grains, resulting in a deficit of small grains in SNRs.

Evidence for dust destruction in SNR shocks is mainly based on different dust mass between that within SNR and that surrounding SNR. \citet{2018ApJ...855...12Z} and \citet{2020A&A...639A..72W} calculated the dust masses of several tens of SNRs in the Milky Way from their extinction and found they were indeed smaller than that of average interstellar dust mass within the same volume. Unfortunately, the estimation of SNR dust mass in the Milky Way suffers serious uncertainty due to highly inhomogeneous distribution of ISM. \citet{2015ApJ...799...50L} presented a systematic study of the complete sample of 61 SNRs in the LMC and derived that a supernova would remove 3.7\,$M_\odot$ dust on average of 12 SNRs with relatively reliable measurements of infrared emission. However, LMC is apparently different from the Milky Way such as in metallicity or galaxy type and subsequently different stellar populations. It may not be able to extend to the impact of supernova explosions in the Milky Way on the dust, but we can refer to the method it used. M31 is very similar to the Milky Way in both metallicity and type, which is the largest external spiral galaxy in the local group and classified as an SA(s)b spiral \citep{1991rc3..book.....D}. The high inclination of $i \simeq$78$^{\circ}$ of M31 raises the surface brightness, and increases the contrast with foreground and background emission. At a distance of d = 744\,kpc \citep{2010A&A...509A..70V}, M31 provides an opportunity to study the SNRs with spatial resolution that is surpassed only by the Magellanic Clouds.

Previous studies of far-infrared (FIR) emission from the dust in M31 include maps made with Infrared Astronomical Satellite\citep[IRAS;][]{1984ApJ...278L..59H,1994AJ....108.1667D}, the Infrared Space Observatory \citep[ISO;][]{1998A&A...338L..33H}, $Spitzer$ \citep{2006ApJ...638L..87G}, and $Herschel$ \citep{2012A&A...546A..34F}.

%
%

The organization of this paper is as follows: the observational data and data reduction are described in Section ~\ref{2}, the dust model and results of SED curve fitting are followed in Section ~\ref{3}, where we estimate and discuss the mass and temperature of SNRs. The results are summarized in Section ~\ref{4}.

\section{Data and Method}\label{2}

\subsection{The sample of supernova remnants in M31}

Identification of SNRs in extra-galaxies is limited by sensitivity and resolution. The established criterion is the spectral line intensity ratio of H$\rm \alpha$ to [$\rm S\,_{\rm II}$]. The early search found a couple of tens SNRs in M31 by e.g. \citet{1981AJ.....86..989D,1981ApJ...247..879B}. Our initial sample consists of the 156 SNR candidates identified by \citet{2014ApJ...786..130L}. They identified the SNRs based on H$\rm \alpha$ and [$\rm S\,_{\rm II}$] images with the criteria on the line ratio of [$\rm S\,_{\rm II}$]/ $\rm H \rm \alpha > 0.4$ and the circular shape. Seventy-six of them were newly found. This is the largest reliable sample of SNRs in M31. According to their analysis, 23$\%$ are Type Ia SNRs candidates and 77$\%$ are core-collapse(CC) SNRs, and their size distributes from a few to about 100 pc with the peak around 40-50\,pc.


\subsection{Infrared image and photometry}

The infrared spectral energy distribution is the basis for estimating the dust mass of SNRs. The present analysis uses the data from $Spitzer$/IRAC by \citet{2006ApJ...650L..45B}, $Spitzer$/MIPS data by \citet{2006ApJ...638L..87G} and  $Herschel$/PACS by the $Herschel$ Exploitation of Local Galaxy Andromeda project \citep[HELGA;][]{2012A&A...546A..34F}. We refer to the images by the camera name and nominal wavelength in microns: IRAC3.6, IRAC4.5, IRAC5.8, IRAC8.0, MIPS24, MIPS70, PACS70, PACS100, PACS160, SPIRE250 and SPIRE350. The FWHM of the IRAC image is 1.6$''$, 1.6$''$, 1.8$''$, 1.9$''$ at 3.6, 4.5, 5.8, and 8\,$\mu$m respectively, while of the MIPS is 6.4$''$, 19.7$''$, 38.7$''$ at 24, 70, 160\,$\mu$m respectively. The FWHM of the $Herschel$ image is 7.7$''$, 12.5$''$, 13.3$''$, 18$''$, 25$''$ at 70, 100, 160, 250, 350\,$\mu$m respectively. Because the cold dust in SNRs is orders-of-magnitude more massive than the hot dust (see e.g. \citet{2003ApJ...597L..33M}), the focus is on the far-infrared (FIR) images, in particular the $Herschel$ images. Due to the limitation of resolution, we select only the objects with the diameter $D > 28$\,pc, i.e. an angular size of $\geq7.7''$ at 744\,kpc, the distance to M31, corresponding to the FWHM of the $Herschel$ 70\,$\mu$m image. For some SNRs with obvious infrared features, the condition is relaxed to 24pc. The whole sample consists of 23 SNRs listed in Table~\ref{table1} with the target series number, position, size and SN type form \citet{2014ApJ...786..130L}.

\begin{figure}
	\centering
	\centerline{\includegraphics[scale=0.3]{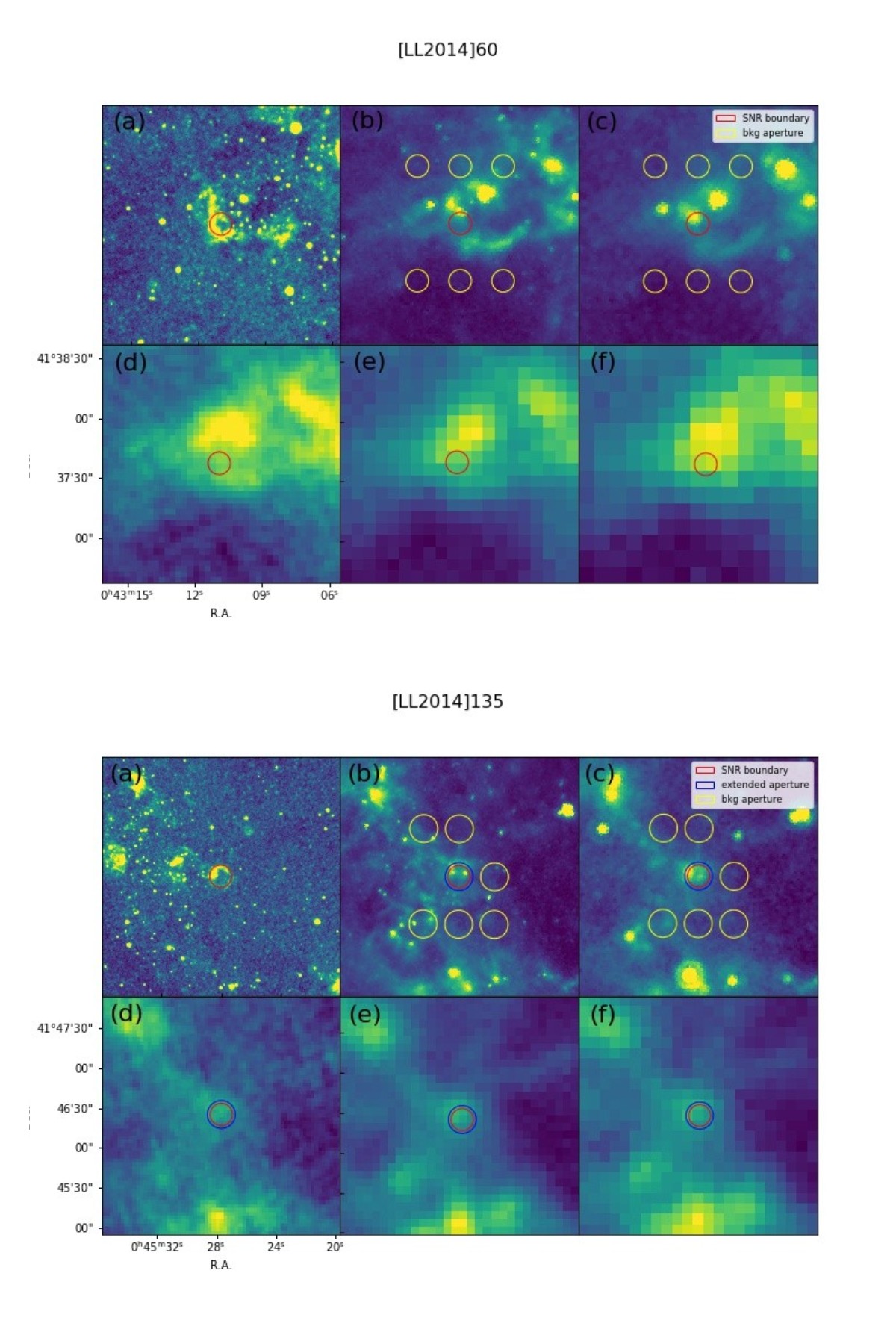}}
	\caption{Images of SNR [LL2014]60 and [LL2014]135 in various bands :(a) $H\alpha$, (b) $Spitzer$/IRAC8, (c) $Spitzer$/MIPS24, (d) Hershcel/PACS70 , (e) $Herschel$/PACS100, (f) $Herschel$/PACS160, where the object is at the center. The red circle marks the area of the SNR, and the six equal-area circles at beside marks the areas used for background evaluation. The blue circle of [LL2014]135(bottom) marks the extended area of the SNR because of the obvious infrared feature.
		\label{fig1}}
\end{figure}

To measure the flux of supernova remnants as extended sources, the aperture photometry is performed. The range of SNRs is taken from \citet{2014ApJ...786..130L}, meanwhile the range is appropriately extended for a few sources whose infrared emission is particularly strong extending obviously beyond the size given by \citet{2014ApJ...786..130L}. Firstly, six circles of the same size around each supernova remnant are selected as the background. The circles are located beside the object with the distance of 30$''$ or 30$\sqrt{2}$ $''$ from the supernova remnant as shown in Figure~\ref{fig1}. For an extended source, it is impossible to include all the flux within a finite aperture, so certain corrections are made to the aperture photometry results according to the flux-aperture correction curve from instrument handbook of $Spitzer$/MIPS\footnote{https://irsa.ipac.caltech.edu/data/SPITZER/docs/mips/mipsinstrumenthandbook/} and $Herschel$/PACS\footnote{http://Herschel.esac.esa.int/hcss-doc-15.0/}.

Uncertainty of photometric results comes from four sources: (1) Noise and flux integration effects related to the source aperture are estimated by $\Delta ap$. This aperture uncertainty is calculated empirically as the standard deviation of the sums from each of the background apertures. (2) The determination of the background level contributes an uncertainty $\Delta bg$, calculated as the standard deviation of all background values, multiplied by a ratio of pixels in the source and background apertures, $N_{\rm ap}/\sqrt{N_{\rm bg}}$. (3) The effect of individual uncertainty in each pixel's flux density from data reduction, summed in quadrature, is $\Delta pix$. The first three are non-systematic uncertainty, and (4) the fourth is instruments' system calibration uncertainty $\Delta cal$. In total, the uncertainty is $\Delta flux^{2}= \Delta ap^{2} +\Delta bg^{2} +\Delta pix^{2}+ \Delta cal^{2}$. The measured fluxes of SNRs are listed in Table~\ref{table1} with the uncertainty.

The relative error at the $Spitzer$/IRAC8.0 and MIPS24 is about 7-8$\%$ and 10$\%$ respectively, which is comparable to 5$\%$ and 10$\%$ given by \citet{2006ApJ...638L..87G}. It can be seen that the relative error at the PACS band is about 12$\%$ at 70 and 100\,$\mu$m and 20$\%$ at 160\,$\mu$m, while at SPIRE250 and SPIRE350, the relative error reaches up to 20-30$\%$. This is apparently higher than the 10$\%$ error for PACS and 7$\%$ for SPIRE by \citet{2012A&A...546A..34F}. Though the error of the \citet{2012A&A...546A..34F} work is dominated by the calibration uncertainty, the error of this work mainly comes from $\Delta ap$ and $\Delta bg$ for the SNRs -- the extended objects.


\section{Results}\label{3}

\subsection{The Dust Model}

The radiation flux $F_{\nu}$ is related to the dust mass $M_{d}$ by the following equation:
$$F_{\nu} = \frac{M_{d}B_{\nu}(T_{d})\kappa_{\nu}(a)}{d^{2}} $$
where $\kappa_{\nu}(a)$, the dust mass absorption coefficient of a spherical particle is
$$\kappa_{\nu}(a) = (\frac{3}{4\pi \rho a^{3}})(\pi a^{2}Q_{\nu}(a)).$$
In addition, $T_{d}$ is the dust temperature, $Q_{\nu}(a)$ is the dust emission efficiency, $a$ and $\rho$ are the dust radius and density respectively.
\citet{2014ApJ...780..172D} calculated $\kappa_{\nu}$ of different dust components and yielded an approximate ratio of amorphous silicate to carbonaceous as 3:1 for the dust in M31. We follow this model: \\
$$\kappa_{\nu}(a) = 0.25\kappa_{\nu}^{\rm carbon}(a)+0.75\kappa_{\nu}^{\rm silicate}(a)$$
The absorption coefficient is calculated with the optical constant of silicate and graphite from\defcitealias{1984ApJ...285...89D}{DL84} \citetalias{1984ApJ...285...89D} \citep{1984ApJ...285...89D} at a constant radius $a=0.1\,\mu m$. The dust size has no effect on the mass opacity coefficient, $\kappa_{\nu}(a)$ in the far-infrared band, as \citet{1983QJRAS..24..267H} already showed that the kappa of spherical dust has nothing to do with the dust size in mid-IR and far-IR based on the Mie theory. Though the grain destruction mechanisms in SNRs will remove the smaller grains from the mixture, if shattering is dominant, the effect of the smaller grain's deficit on the opacity coefficient exists mainly in visual or violet rather than in mid-IR and far-IR. This can be understood since the infrared wavelengths adopted here starting from 70\,$\mu$m are orders of magnitude larger than the dust size. Therefore, assuming one-size dust would not influence the result.

The flux at 24\,$\mu$m and shorter wavelength is very much influenced by spectral line emission and stellar radiation, they are not taken into account in fitting. Besides, the low resolution of $Herschel$/SPIRE leads to only an upper limit of the SNR's flux at 250\,$\mu$m and 350\,$\mu$m. As a result, the reliable measurements are available at 70, 100 and 160\,$\mu$m. Therefore, only one-temperature component is assumed for fitting the SED, which prefers cold dust. Though warm dust may be present, its mass should be significantly smaller than the cold dust and negligible. In an experimental running with two dust components,  it is found that the warm dust accounts for less than $1\%$ of total dust mass, which agrees with other studies, e.g. \citet{2012MNRAS.420.3557G} on the Tycho and Kepler SNR and  \citet{2003Natur.424..285D} on Cas A. This can be explained by very high sensitivity of dust mass to temperature. \citet{2015ApJ...799...50L} also take only one component for a non-linear least square fitting to the SEDs of SNRs in LMC.

\begin{figure}
	\centering
	\centerline{\includegraphics[scale=1]{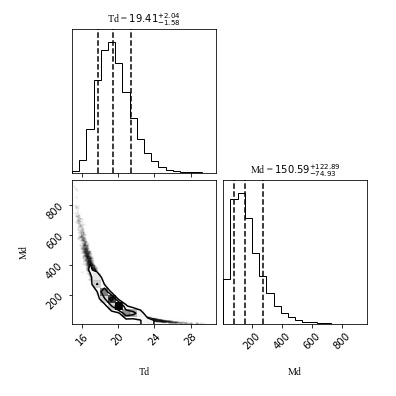}}
	\caption{The corner plot of  the fitting to the infrared spectral energy distribution of SNR [LL2014]12 with 20,000 simulations, and the histograms show the distribution of dust temperature and  mass.
		\label{fig2}}
\end{figure}

The errors of mass and temperature are estimated using Bootstrap that is a Nonparametric Monte Carlo method. Bootstrap is done by randomly sampling within Gaussian error and fitting the resultant SED. Figure~\ref{fig2} presents the case of the SNR[LL2014]12 as an example to show the distribution of dust mass and temperature with 20,000 simulated data sets. The error of the parameters for all SNRs are listed in Table~\ref{table1}.

\subsection{Results}

\begin{figure}
	\centering
	\centerline{\includegraphics[scale=0.45]{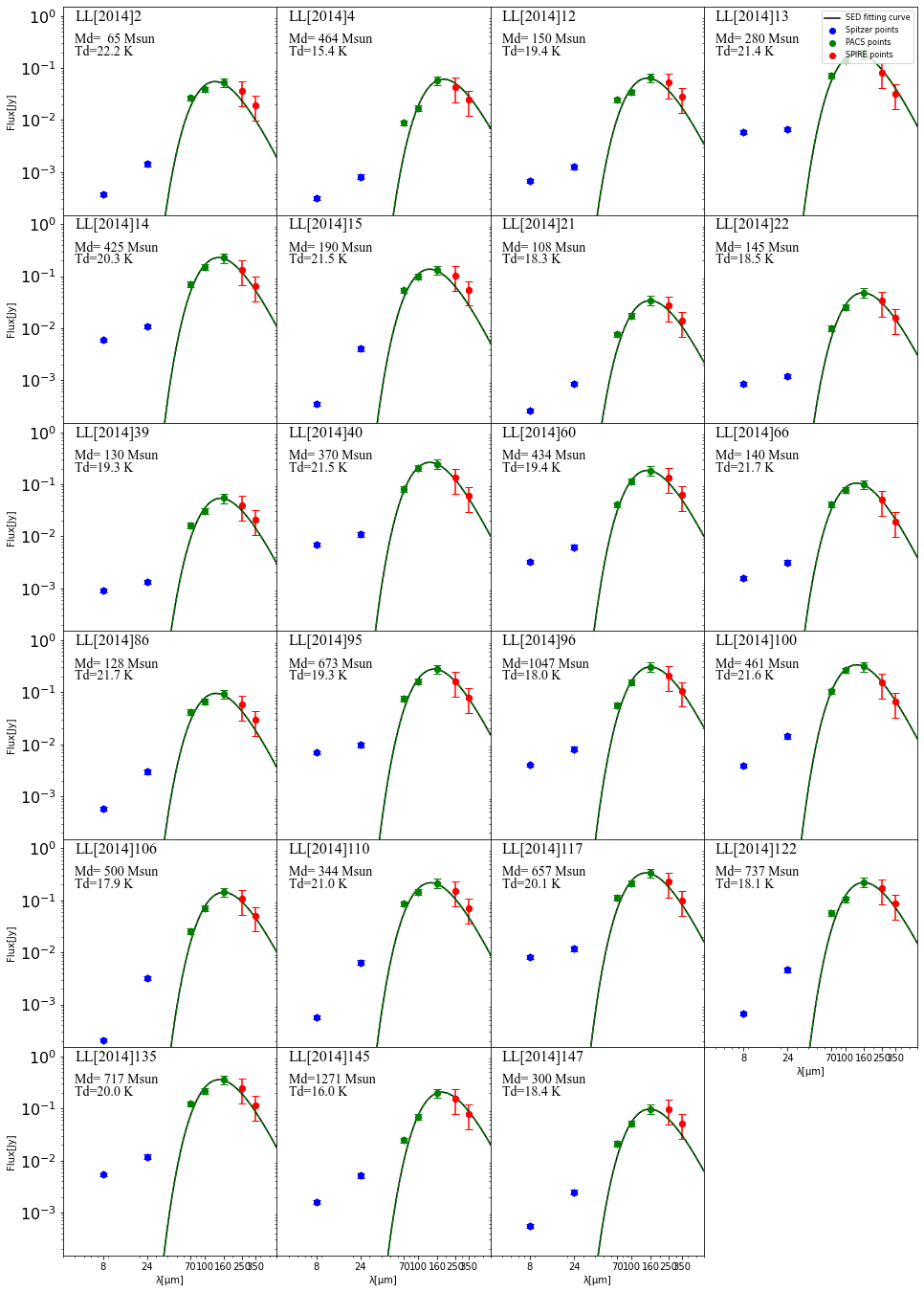}}
	\caption{The fitting results to the SED with the derived dust temperature and mass. The blue, green and red dots denote the photometry results from $Spitzer$, $Herschel$/PACS and $Herschel$/SPIRE respectively, where both the blue and red dots are not taken into account in fitting while the red dots are taken as a reference of upper limit at corresponding wavelengths.
		\label{fig3}}
\end{figure}

The fitting results are displayed in Figure~\ref{fig3} together with the fluxes estimated from the $Spitzer$ and $Herschel$ observation with their errors. The fitted parameters, i.e. dust mass $M_{\rm d}$ and temperature $T_{\rm d}$, are listed in Table~\ref{table1} with the errors. The spatial distribution shows that the studied SNRs are concentrated around the ring with R=11.2\,kpc where photometry is better performed.  

\subsubsection{The Dust Temperature}

The temperature of dust is expected to be higher in the SNRs due to the heating by SN explosion than the general molecular cloud, thus $T_d$ can diagnose whether we are probing the dust in SNRs. \citet{2014ApJ...780..172D} calculate the large-scale dust temperature distribution that yields $T_{\rm d} \sim$17\,K around the 11.2\,kpc ring. Among the 23 objects, only two objects [LL2014]4 and [LL2014]145 have $T_{\rm d} < $17\,K being 15.4\,K and 16.0\,K respectively, one more object [LL2014]106 has $T_{\rm d} = 17.9$\,K very close to 17\,K. The fact that the dust temperature is very close to the average interstellar dust means that these SNRs have cooled down and completely mixed with the interstellar medium, thus they are removed from the following analysis of the SNR dust mass. Excluding the three objects, $T_{\rm d}$ of other SNRs is then $20.1^{+1.8}_{-1.5}$, higher than the average $T_{\rm d}$ of the ring. This is consistent with the heated dust of SNRs. Although the SNRs we selected are relatively large in radius, which means that they are old, it can still be seen that the dust heated by the SN explosion has not completely cooled down. 
In comparison, the average $T_{\rm d}$ of SNRs is found to be 23.3\,K in LMC by \citet{2015ApJ...799...50L}, slightly higher than the case in M31. This can be understood from its relatively old age of the SNRs sample in M31 because of its order of magnitude large distance and consequently only large ones resolved at far-infrared wavelengths.

\subsubsection{The Dust Mass}

The dust mass  ranges from about 100 to 800\,$ M_{\odot}$, with a peak around 205\,$ M_{\odot}$. This looks huge as a total amount of dust in comparison with several to tens of solar mass for a supernova remnant in the Milky Way \citep{2020A&A...639A..72W} with comparable metallicity.  However, the radius of these SNRs shown in Table~\ref{table1} is generally bigger than 30\,pc and up to about 80\,pc, indicating that they have passed the Sedov and even the snowplow phase. On the other hand, the SNRs studied by \citet{2020A&A...639A..72W} are mostly in the Sedov phase with a radius smaller than 20\,pc. Considering that the dust mass is proportional to the radius cubed, the order of magnitude difference is acceptable, in particular if the slightly higher metallicity of M31 is considered.  This result may reflect the dust mass of a supernova remnant during the snowplow phase.

In spite of the huge total dust mass, the average dust surface density is $0.10^{+0.07}_{-0.04}$\,${M_{\odot} \rm pc^{-2}}$, which is about half of $\sim$ 0.22\,${M_{\odot} \rm pc^{-2}}$, the density around the 11.2\,kpc ring in M31 derived by \citet{2014ApJ...780..172D} where most of the studied SNRs locate. This difference means about $\sim50\%$ dust is destroyed by the supernova explosion, which agrees with the result on the Monoceros SNR by \citet{2018ApJ...855...12Z}.  Similarly,  the dust surface density is derived to be $<$0.01\,${ M_{\odot} \rm pc^{-2}}$ for the SNRs in LMC by \citet{2015ApJ...799...50L}, about a few tenths of the surrounding density, which agrees with the case in M31 as well. Meanwhile, the difference in the amount of dust may be partly caused by the metallicity, and the range of SNR in LMC is smaller on average so that the dust is comparatively in the inner area and more destroyed by the SN explosion, but the SNRs in M31 may contain more dust in the outer area including some interstellar dust unaffected by SN explosion. It should be noted that there is substantial evidence that the \citet{2014ApJ...780..172D} estimates of dust surface density in M31 are too large, by a factor of $\sim$2.5 in comparison with \citet{2015ApJ...814....3D} and \citet{2019MNRAS.489.5436W}. This means that both the surrounding dust surface density and the SNR dust mass could be decreased by a factor of $\sim$2.5 if using other values of opacity, but it will not change the fraction of dust destroyed because both are inversely proportional to the dust opacity. On the other hand, it will reduce the dust mass of SNRs in M31 by a factor of $\sim$2.5, which then brings about a closer agreement in the dust mass with the Galactic SNRs and in the dust surface density with the LMC SNRs.

\section{Summary}\label{4}

With the infrared images at 70, 100 and 160\,$\mu$m by the $Herschel$ HELGA survey, and at the $Spitzer$/IRAC4 and MIPS24\,$\mu$m, the photometry is performed to the supernova remnants in M31 identified from their [SII] and H$\alpha$ emission ratio by \citet{2014ApJ...786..130L}. Based on the infrared spectral energy distribution resulted from the photometry, the dust mass and temperature of 23 SNRs with relatively reliable photometry  is estimated by a one-temperature dust model fitting with the dust parameters derived from the overall infrared emission of M31 by \citet{2014ApJ...780..172D}, meanwhile the uncertainty is calculated by bootstrap.

The average dust temperature of 20 SNRs is found to be $20.1^{+1.8}_{-1.5}$\,K, which is higher than the temperature of the dust at 11.2\,kpc \citep{2014ApJ...780..172D} where most of the SNRs locate, indicating the heating effect by the supernova explosion. The other three SNRs exhibit no obvious increase in the dust temperature, which may be caused by their very old age. For the relatively warm 20 SNRs, the dust mass is estimated to range from about 100 to 800\,$ M_{\odot}$, much higher than that in the Milky Way, which can be partly accounted for by the large size of several tens of parsec and possibly also by a small dust absorption coefficient. On the other hand, the dust surface density, $0.10^{+0.07}_{-0.04}$\,${ M_{\odot} \rm pc^{-2}}$, is about half of the surrounding area, which agrees with the case of SNRs in LMC found by \citet{2015ApJ...799...50L}, and implies that about half of the dust in the SNR area is destroyed by the SN explosion. The dust temperature, the radius, and thus the dust mass all demonstrate that the studied SNRs are old and in the snowplow or even fadeaway phase due to the far distance of M31 and the result can serve as a reference for the final effect of SN explosion on the surrounding dust.

%

\acknowledgments{We thank Profs. Maohai Huang, Chaowei Tsai and Jian Gao for their help and discussion, and the referee for very careful reading and helpful suggestions. This work is supported by the NSFC projects 12133002 and 11533002, National Key R\&D Program of China No. 2019YFA0405503 and CMS-CSST-2021-A09. This work made use of the data taken by \emph{$Herschel$} and \emph{$Spitzer$}.
}

\bibliographystyle{aasjournal}
\bibliography{ccm2}

\begin{sidewaystable}
	\begin{center}
		\caption{\label{table1}The fundamental information and derived dust parameters of 23 supernova remnants in M31}
		\vspace{0.05in}
		\setlength{\tabcolsep}{1.0mm}
		\begin{tabular}{lcccrrrrrrrrrrr}
			\tableline
			\multirow{2}{*}{Name}{\tablenotemark{a}} &$\rm R.A.${\tablenotemark{a}}& $\rm Decl${\tablenotemark{a}} & $D${\tablenotemark{a}} & $\rm Type${\tablenotemark{a}} &$F_{\rm 24\mu m}$&$\sigma_{\rm 24\mu m}$ &$F_{\rm 70\mu m}$&$\sigma_{\rm 70\mu m}$ &$F_{\rm 100\mu m}$ &$\sigma_{\rm 100\mu m}$&$F_{\rm 160\mu m}$ &$\sigma_{\rm 160\mu m}$ &$M_{\rm d}$ & $T_{\rm d}$ \\
			&(h m s) & ($^{\circ}$ $'$ $''$) & (pc) & & (mJy)&(mJy) &(mJy) &(mJy)& (mJy)&(mJy) &(mJy) &(mJy) &($ M_{\odot} $) & (K) \\
			\tableline
			[LL2014]2 &00 39 23.3&40 44 19.8&55.4&CC& 1.1 &0.09& 26.6 &3.2&38.8 &4.7&53.2&6.4&$66^{+47.5}_{-28.8}$&$22.2^{+2.1}_{-1.8}$\\
			\tableline
			[LL2014]4 &00 39 44.9&40 29 54.1&45.6&CC&0.3&0.02&8.9&1.1& 16.9&2.0&57.7&6.9&$464^{+359.0}_{-224.4}$&$15.4^{+1.3}_{-1.0}$\\
			\tableline
			[LL2014]12 &00 40 32.0&40 30 31.4&51.2&CC&0.8&0.06&24.4&2.9&34.2&4.1&65.7&7.9&$150^{+122.9}_{-74.9}$&$19.4^{+2.0}_{-1.6}$\\
			\tableline
			[LL2014]13 &00 40 33.4&40 43 36.0&55.2&CC&4.2&0.34&71.2&8.5&140.1&16.8&186.6&22.4&$280^{+193.3}_{-112.6}$&$21.4^{+1.7}_{-1.6}$\\
			\tableline
			[LL2014]14 &00 40 33.6&40 32 47.1&59.6&CC&9.4&0.75&70.2&8.4&150.4&18.1&225.1&27.0&$425^{+288.2}_{-174.7}$&$20.3^{+1.7}_{-1.4}$\\
			\tableline
			[LL2014]15 &00 40 36.1&40 49 09.9&63.6&CC&2.1&0.17&53.5&6.4&98.1&11.8&132.2&15.8&$190^{+128.3}_{-80.5}$&$21.5^{+1.9}_{-1.6}$\\
			\tableline
			[LL2014]21 &00 41 03.9&40 47 02.2&28.0&Ia&0.2&0.02&7.6&0.9&17.2&2.1&34.3&4.1&$108^{+78.4}_{-50.4}$&$18.3^{+1.6}_{-1.3}$\\
			\tableline
			[LL2014]22 &00 41 06.2&40 52 08.8&37.6&Ia&0.8&0.06&9.9&1.2&25.7&3.1&48.4&5.8&$145^{+97.8}_{-62.8}$&$18.5^{+1.4}_{-1.3}$\\
			\tableline
			[LL2014]39 &00 42 14.8&40 52 03.0&32.4&Ia&2.2&0.18&16.2&1.9&30.5&3.7&54.6&6.6&$131^{+88.8}_{-58.8}$&$19.3^{+1.7}_{-1.4}$\\
			\tableline
			[LL2014]40 &00 42 20.2&41 27 54.8&45.2&CC&9.3&0.74&80.7&9.7&205.1&24.6&246.6&29.6&$371^{+219.2}_{-135.8}$&$21.5^{+1.6}_{-1.5}$\\
			\tableline
			[LL2014]60 &00 43 10.9&41 37 38.3&45.2&CC&5.2&0.42&40.4&4.9&115.2&13.8&184.3&22.1&$434^{+281.8}_{-185.0}$&$19.4^{+1.5}_{-1.3}$\\
			\tableline
			[LL2014]66 &00 43 27.9&41 18 30.5&40.4&CC&1.8&0.14&41.5&5.0&77.2&9.3&100.9&12.1&$140^{+93.6}_{-58.7}$&$21.7^{+1.9}_{-1.6}$\\
			\tableline
			[LL2014]86 &00 44 13.2&41 50 30.7&44.6&Ia&1.6&0.13&42.6&5.1&66.3&8.0&93.5&11.2&$128^{+89.9}_{-56.5}$&$21.7^{+2.0}_{-1.6}$\\
			\tableline
			[LL2014]95 &00 44 29.5&41 52 44.1&55.0&CC&10.2&0.82&75.4&9.1&165.1&19.8&281.8&33.8&$673^{+461.7}_{-306.8}$&$19.3^{+1.7}_{-1.4}$\\
			\tableline
			[LL2014]96 &00 44 32.1&41 23 35.9&54.8&CC&4.7&0.38&59.9&7.2&155.8&18.7&310.1&37.2&$1047^{+733.3}_{-444.6}$&$18.0^{+1.5}_{-1.3}$\\
			\tableline
			[LL2014]100 &00 44 38.8&41 25 26.8&46.2&CC&13.6&1.09&104.8&12.6&266.3&31.9&314.0&37.7&$461^{+256.5}_{-181.7}$&$21.6^{+1.6}_{-1.4}$\\
			\tableline
			[LL2014]106 &00 44 47.0&41 29 20.5&42.0&CC&0.7&0.06&25.9&3.1&71.2&8.5&143.8&17.3&$500^{+348.6}_{-221.4}$&$17.9^{+1.5}_{-1.2}$\\
			\tableline
			[LL2014]110 &00 44 50.8&41 32 12.2&67.8&CC&5.5&0.44&86.7&10.4&146.1&17.5&216.3&25.9&$344^{+245.2}_{-149.6}$&$21.0^{+1.8}_{-1.6}$\\
			\tableline
			[LL2014]117 &00 45 05.0&41 38 53.2	&69.8&CC&10.3&0.82&114.4&13.7&212.1&25.4&338.0&40.6&$657^{+464.5}_{-283.3}$&$20.1^{+1.8}_{-1.5}$\\
			\tableline
			[LL2014]122 &00 45 12.1&41 50 17.5&72.4&Ia&1.1&0.09&58.4&7.0&105.7&12.7&225.8&27.1&$737^{+583.7}_{-375.0}$&$18.1^{+1.8}_{-1.3}$\\
			\tableline
			[LL2014]135 &00 45 27.8&41 46 27.0&67.2&CC&8.1&0.65&123.2&14.8&220.1&26.4&361.4&43.3&$717^{+550.8}_{-344.7}$&$20.0^{+2.0}_{-1.5}$\\
			\tableline
			[LL2014]145 &00 45 56.5&42 11 10.5	&88.8&Ia&3.6&0.29&24.6&3.0&69.5&8.3&200.1&24.0&$1271^{+950.7}_{-567.8}$&$16.0^{+1.2}_{-1.0}$\\
			\tableline
			[LL2014]147 &00 46 20.1&41 52 59.7&89.4&CC&1.8&0.14&21.6&2.6&51.1&6.1&98.7&11.8&$300^{+218.3}_{-142.9}$&$18.4^{+1.7}_{-1.3}$\\	
			\tableline
			\multicolumn{15}{l}{{\large $^a$} from \citet{2014ApJ...786..130L}}
		\end{tabular}
	\end{center}
\end{sidewaystable}

%
%

\end{CJK*}
\end{document}